# Dual-Comb Coherent Raman Spectroscopy with Lasers of 1-GHz Pulse Repetition Frequency

Kathrin J. Mohler,[1,2] Bernhard J. Bohn,[1,2] Ming Yan,[1,2] Theodor W. Hänsch,[1,2] and Nathalie Picqué [1,2,*]

[1]Ludwig-Maximilians-Universität München, Fakultät für Physik, Schellingstrasse 4/III, 80799 München, Germany
[2]Max-Planck-Institut für Quantenoptik, Hans-Kopfermann-Strasse 1, 85748 Garching, Germany
*Corresponding author: nathalie.picque@mpq.mpg.de

**Abstract:**
We extend the technique of multiplex coherent Raman spectroscopy with two femtosecond mode-locked lasers to oscillators of a pulse repetition frequency of 1 GHz. We demonstrate spectra of liquids, which span 1100 cm$^{-1}$ of Raman shifts. At a resolution of 6 cm$^{-1}$, their measurement time may be as short as 5 μs for a refresh rate of 2 kHz. The waiting period between acquisitions is improved ten-fold compared to previous experiments with two lasers of 100-MHz repetition frequencies.

Laser frequency combs enable new approaches to molecular spectroscopy and sensing [1-12]. One of these approaches [7] harnesses coherent nonlinear Raman effects in condensed matter with a pump-probe scheme involving two femtosecond combs of slightly different pulse repetition frequencies. Dual-comb coherent anti-Stokes Raman spectroscopy (CARS) was demonstrated with spectra, spanning around 1000 cm$^{-1}$, measured on a μs-ms time scale with a resolution limited by the intrinsic width of the vibrational molecular bands.

Ref. [7] details the principle of dual-comb coherent Raman spectroscopy. In summary, in the time domain (Fig.1), the pulses of one comb excite, with a periodicity of $1/(f_{rep} + \delta f_{rep})$, low-lying vibrational levels in a Raman two-photon process. The refractive index of the sample is modulated at the vibrational period, with a ring-down time related to the coherence time of the transition. After each excitation, the pulse of a second comb of repetition frequency $f_{rep}$ probes the vibrational excitation at a linearly increasing time delay, automatically induced by the small difference in repetition frequencies $\delta f_{rep}$. The spectrum of the probe comb is alternately blue- and red-shifted [13] and its intensity modulation, measured behind an edge spectral filter, directly reveals the frequencies of the excited molecular vibrations with a down–conversion factor equal to $\delta f_{rep}/f_{rep}$. A Fourier transform of such time-domain interference computes the Raman spectrum.

As in other dual-comb techniques, the time separation between two pulses of a pump-probe pair is periodically scanned, by steps of $\delta f_{rep}/f_{rep}^2$, over a range from 0 to $1/f_{rep}$. Therefore, for an optimal use of the experimental time, the vibrational decay time of the molecular sample (set by the coherence time or by the signal-to-noise ratio) should be similar to $1/f_{rep}$. In the frequency-domain picture, the desired spectral resolution should be comparable to the comb line spacing. The ratio of the comb line spacing to the resolution gives therefore the duty cycle of the acquisition sequence. As the coherence time of the molecular bands in the liquid phase is typically of tens of ps, corresponding to line widths of several tens of GHz (several cm$^{-1}$), frequency combs of large line spacing would benefit dual-comb CARS, since the refresh rate $\delta f_{rep}$ of successive interferograms may be increased. However, coherent Raman four-wave mixing signals scale with the cube of the peak power of the laser pulses, making the use of high repetition frequency lasers challenging.

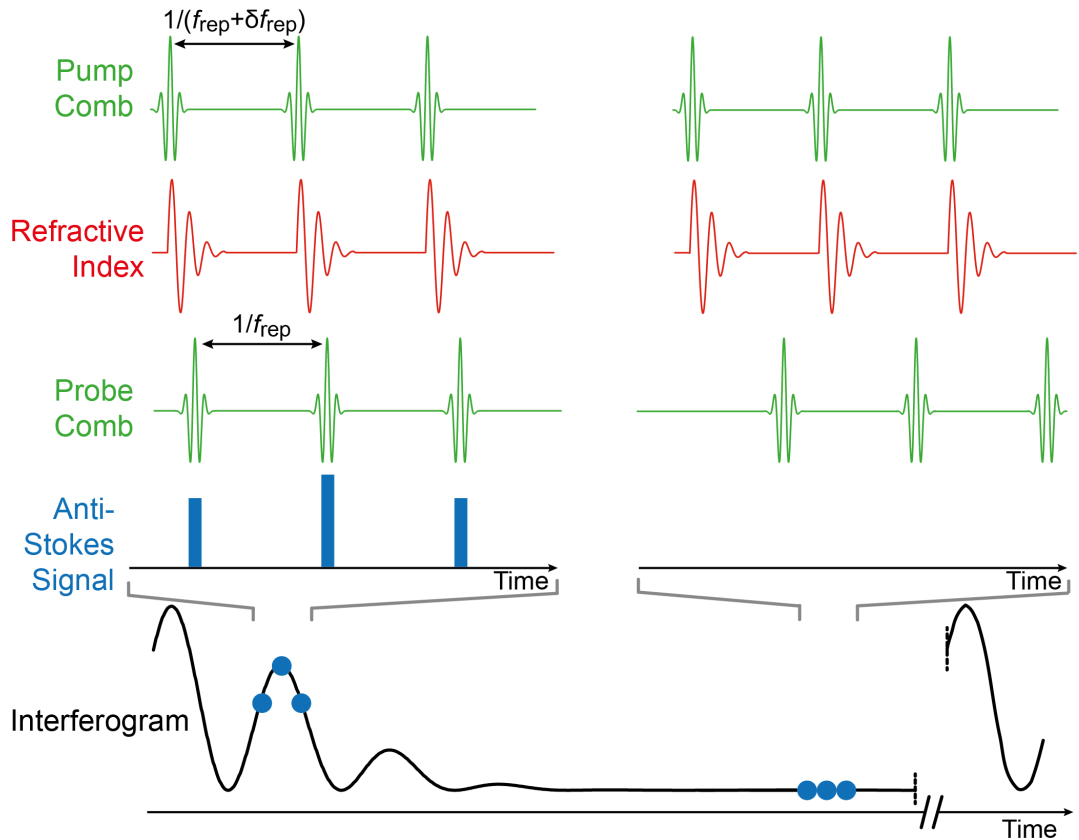

**Figure1:** Time-domain principle of dual-comb CARS. See text and ref. [7] for details.

The first demonstrations [7] of dual-comb CARS have been performed with 100-MHz oscillators of 20-fs pulse duration and the duty cycle was limited to $3 \times 10^{-3}$. In this letter, we investigate the potential of 1-GHz oscillators for dual-comb CARS.

Figure 2 shows the experimental set-up. Two identical titanium-sapphire femtosecond mode-locked lasers (Laser Quantum, Taccor-10s) are used. Their spectrum is centered around 795 nm (12580 cm$^{-1}$) with a full width at half maximum (FWHM) of 60 nm (880 cm$^{-1}$). At the output of the sealed oscillators, the pulses are chirped with a specified group delay dispersion of about +450 fs$^2$. Their repetition frequency is around 1 GHz and the difference in repetition frequencies $\delta f_{rep}$ is chosen between 500 Hz and 2000 Hz. The repetition frequency is controlled by piezo-electric transducers, which adjust the laser cavity length. Low-bandwidth (about 10 kHz) active stabilization is achieved with a servo controller (Laser Quantum, TL-1000), which phase-locks the repetition frequency to an external reference provided by a radio-frequency synthesizer (Rohde & Schwarz, SMA100A-B22).

The two laser beams are linearly polarized with axes that are not perfectly orthogonal. They are combined on a pellicle beam splitter (Thorlabs, CM1-BP145B2). One output of the beam splitter is used for CARS, as detailed below. The second output measures the residual time-domain interference between the two femtosecond lasers on a fast photodiode (Thorlabs, PDA10A): the sequence of bursts at a period of $1/\delta f_{rep}$ serves as a trigger signal for the acquisition of the dual-comb coherent Raman interferograms. On the CARS beam path, chirped mirrors (Layertec) compensate for the group delay dispersion (-940 fs$^2$ in total) in the entire set-up (lasers, optical elements before the focal point, air). Spectral shearing interferometry is employed to characterize the ultrashort pulses by means of a SPIDER device (Venteon, Laser Quantum).

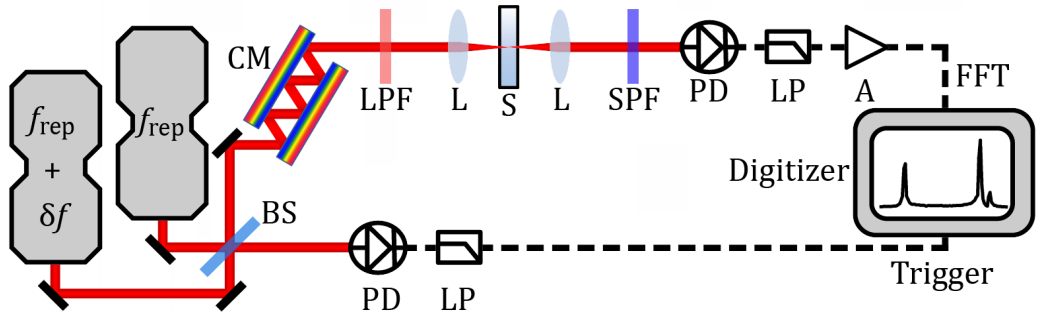

**Figure 2:** Experimental setup for dual-comb CARS using two 1-GHz lasers (see text for details). BS, beamsplitter; CM, chirped mirrors; LPF, long-pass filter; L, lens; S, sample; SPF, short-pass filter; PD, photodetector; LP, electronic long-pass filter; A, amplifier.

A combination of an optical long-wavelength-pass filter (cut-off: 750 nm – 13333 cm$^{-1}$), before the sample, and a short-wavelength-pass filter (cut-off: 750 nm – 13333 cm$^{-1}$), after the sample, separates the anti-Stokes radiation from the light of the titanium-sapphire lasers. The cut-off wavelengths can be slightly changed by tilting the interference filters. The beam is focused, with an aspheric lens of a focal length of 8 mm (Newport, 5724), on the liquid sample contained in a 1 mm–thick quartz cuvette (Starna, Type 21). At the sample, for each laser, the pulse duration is less than 20 fs and the pulse energy 0.5 nJ. The anti-Stokes radiation, emerging from the cuvette in the forward direction, is collimated by a lens identical to the focusing one. It is optically filtered and focused onto a fast silicon photodiode (Newport, 1621). The interference signal is detected together with a large background of other signals originating from unfiltered laser radiation, from nonlinear phenomena (including coherent Raman and non-resonant effects) generated at the sample by each laser and from residual intensity noise. The close-to-orthogonal polarizations of the two comb beams diminishes the interferometric non-resonant background, while retaining a significant part of the resonant contribution to the interference thanks to the depolarization of the sample. Moreover, the non-interferometric signals mostly occur at the repetition frequency of the lasers and its harmonics. They can therefore be electrically filtered out. After filtering (Mini Circuits, ZFHP-0R055-S+ and BLP-200+) and amplification (L-3 Narda-MITEQ, AU-1332), the time-domain interference signal is recorded by a data acquisition board (AlazarTech, ATS9360) with a rate of $10^9$ samples s$^{-1}$.

The interferograms are processed by an automated custom MATLAB program. As in other Fourier transform CARS techniques [14-17], the interferometric non-resonant background is only present at short optical delays. It is therefore entirely suppressed by suitably positioning the numerical triangular apodization window. The length of the apodization window sets the resolution and 6-fold zero-filling of the interferogram interpolates the spectrum. As the two 1-GHz lasers have similar pulse energies at the sample, each laser is alternately the pump and the probe and the interferograms are symmetric about zero optical delays. For instance, the duty cycle, defined in [7] as the ratio 3.6 $f_{rep}/ \delta\nu$ where $\delta\nu$ is the optical resolution, is 2% for an apodized resolution of 6 cm$^{-1}$. This is ten-fold higher than with 100-MHz systems. Here however, only one side of the interferogram about the zero optical delay is Fourier transformed, bringing the actual duty cycle to 1%.

The Fourier transform reveals radio-frequency spectra. Their free spectral range varies from 21 MHz ($\delta f_{rep}$ = 500 Hz, down-conversion factor: 5 10$^{-7}$) to 84 MHz ($\delta f_{rep}$ = 2000 Hz, down-conversion factor: 2 10$^{-6}$). The radio-frequency x-scale is *a posteriori* converted to a scale of optical Raman-shifts by division by the down-conversion factor, with the help of the experimentally measured repetition frequencies. The span of Raman shifts, converted to wavenumbers, is 300-1400 cm$^{-1}$. It is limited on the low-wavenumber side by the optical filters and on the high-wavenumber side by the spectral bandwidth of the lasers. When spectra are averaged, the wavenumber axis of each individual spectrum is corrected by a multiplying factor prior to averaging. Such multiplying factor is determined by assigning one or more line positions to their mean experimental values. This compensates for the residual fluctuations of the difference in repetition frequencies.

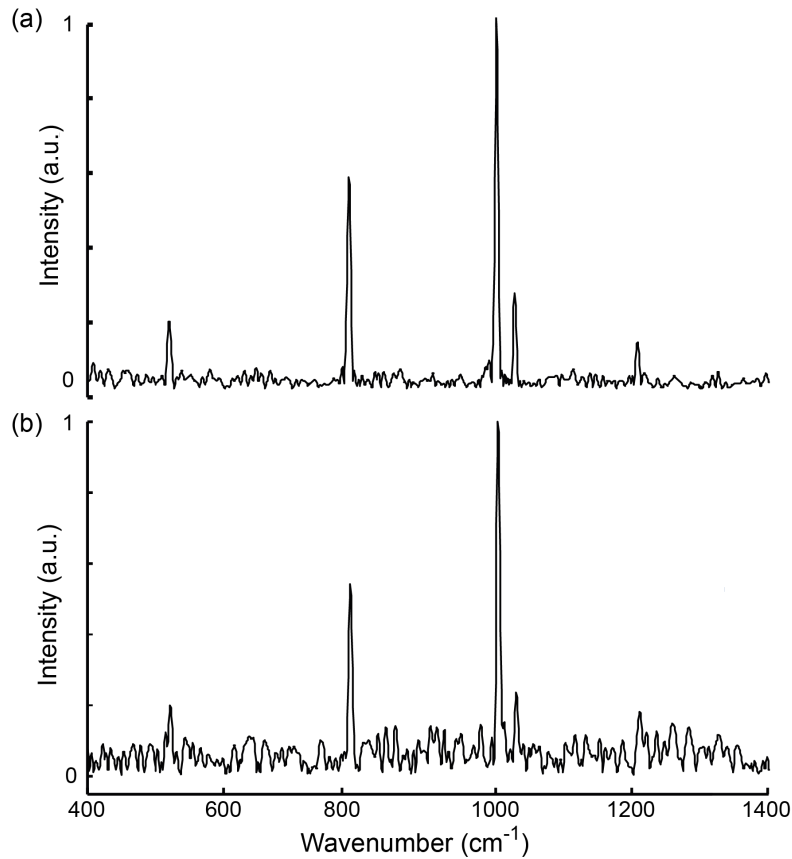

**Figure 3:** Experimental spectra of toluene measured at 6-cm$^{-1}$ resolution within a) 20 μs and b) 5 μs.

Figure 3 shows two experimental dual-comb CARS spectra of neat toluene measured within times on the microsecond scale. No averaging is performed. The apodized resolution is 6 cm$^{-1}$. Therefore the spectra comprise 183 spectral elements (defined as span divided by resolution). The duty cycle is 1% if one only considers single-sided interferograms. Five intense lines of toluene, centered at 522 cm$^{-1}$, 788 cm$^{-1}$, 1001 cm$^{-1}$, 1028 cm$^{-1}$ and 1210 cm$^{-1}$, respectively, are assigned [18] to the $\nu_{13}$, $\nu_{12}$, $\nu_{11}$, $\nu_{10}$, $\nu_8$ bands, respectively. In Fig. 3a, the difference in pulse repetition frequencies is $\delta f_{rep}$ = 500 Hz, which leads to a measurement time of 20 μs. The signal-to-noise ratio for the strongest transition at 1001 cm$^{-1}$ is 110. The noise is taken as the standard deviation of the baseline in the wavenumber region around 1350 cm$^{-1}$. Double-sided interferograms repeat at a refresh time of 1/ $\delta f_{rep}$ =2 ms. Every 2 ms, it is possible to measure two spectra, like that shown in Fig.3a, within 20 μs and with a time separation of a couple of μs, set by the apodization window. In Fig. 3b, as $\delta f_{rep}$ = 2000 Hz, the measurement time is 5 μs, the signal-to-noise ratio is 30 and the refresh time is 500 μs.

Figure 4 shows spectra of mixtures of toluene and 2-propanol at an apodized resolution of 6 cm$^{-1}$. The difference in repetition frequencies is $\delta f_{rep}$ = 500 Hz, and the displayed spectra are the result of one hundred averages. The measurement time is 2 ms and the total experimental time, which includes the dead time due to the mismatch between the repetition frequency and the spectral resolution, is 200 ms. Panels a,b,c,d, respectively, of Fig. 4 present the spectra for mixing ratios of toluene of 100%, 80%, 60% and 40%, respectively. The signal-to-noise ratio of the most intense toluene band is 840, 610, 490 and 280 respectively. The transition at 816 cm$^{-1}$, that one can distinguish in Fig.4c,4d corresponds to the excitation of the symmetric C-C-O stretching mode in 2-propanol [19].

The maxima of the toluene band profiles are found to decrease linearly with a decreasing concentration. With coherent Raman signals, a quadratic concentration dependence is expected. The observed linear behavior implies that the detected signal results from the interference of the CARS electric field with another field [20,21]. Moreover, the line maxima are found (not shown in the figure) to vary linearly with the energy of the probe pulses. Therefore, the local oscillator in this additional interferometric mixing process is the probe beam. The amount of spectral overlap at the detector between the mutually coherent probe and interferometric CARS signals is experimentally controlled by empirically adjusting the cut-off wavelengths of the optical filters placed before and after the sample. The probe pulses always coincide in time with the interferometric CARS signal at the detector. Such heterodyne detection of the weak CARS signal improves the signal-to-noise ratio and gives a linear concentration dependence.

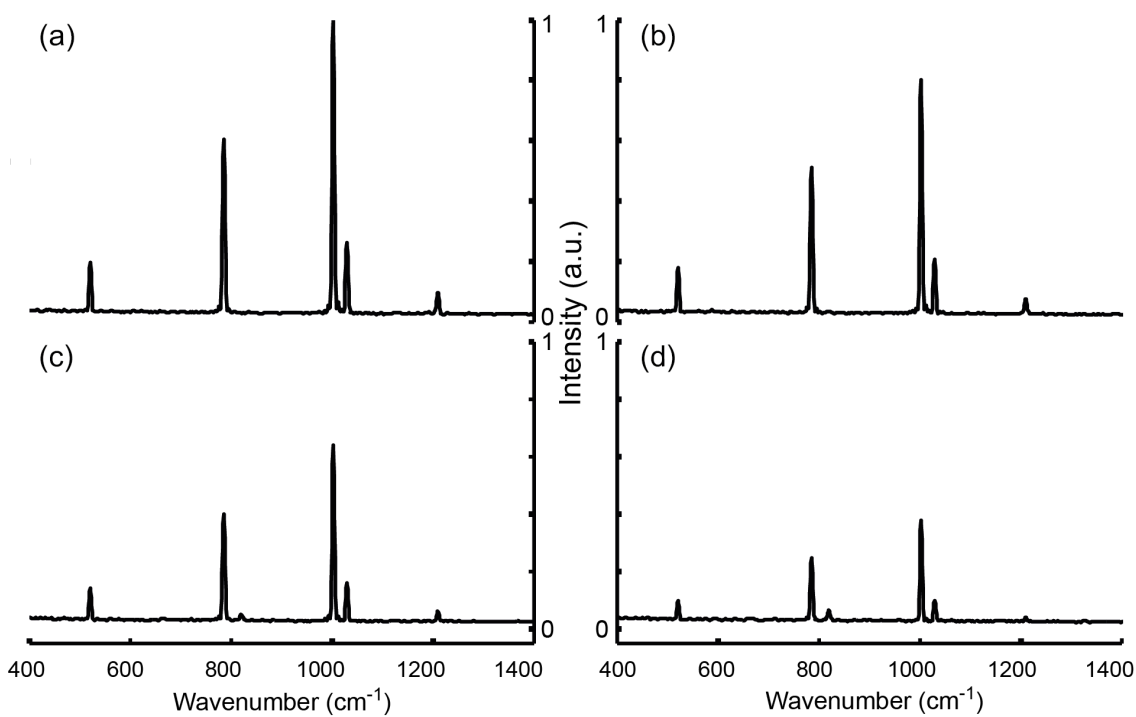

**Figure 4:** Dual-comb Raman spectra of mixtures of toluene and 2-propanol with a) 100%, b) 80%, c) 60%, d) 40% of toluene.

The spectra shown in Figure 5 are that from a) neat toluene, b) neat cyclohexane (band $\nu_5$ at 802 cm$^{-1}$ [22]), c) neat chloroform (band $\nu_3$ at 366 cm$^{-1}$ and band $\nu_2$ at 678 cm$^{-1}$ [23]) and d) a mixture of toluene, cyclohexane and chloroform (mixing ratio of 1:1:1). The difference in repetition frequencies is $\delta f_{rep}$ = 600 Hz. The apodized resolution is 6 cm$^{-1}$ and 120 spectra are averaged in each trace. The measurement time is 2 ms and the total experimental time 200 ms. Figure 5d illustrates both our broad spectral span - with bands centered at 386 cm$^{-1}$ and at 1210 cm$^{-1}$ - and our resolution, good enough to separate the neighboring transitions of toluene at 788 cm$^{-1}$ and if cyclohexane at 802 cm$^{-1}$.

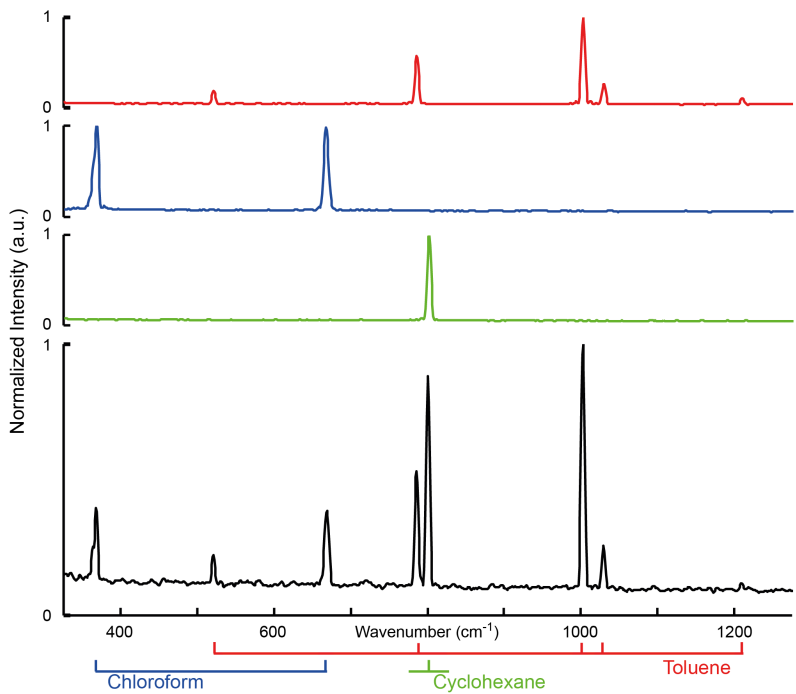

**Figure 5:** Dual-comb CARS spectrum of a) neat toluene, b) neat chloroform, c) neat cyclohexane, d) toluene, chloroform and cyclohexane in mixing ratio 1:1:1.

Because the CARS signal scales with the cube of the peak power, the main difficulty associated to the use of laser systems of high repetition frequency is related to the reduced pulse energy at the laser output. We compare the spectrum of toluene of Figure 3a to the same spectrum (Fig.6) measured with the set-up described in [7]. It uses two 100-MHz titanium-sapphire mode-locked lasers, which emit pulses of 20-fs duration. In the spectrum of Fig. 6, the pulse energy at the sample is 4 nJ per laser, the difference in repetition frequencies is 50 Hz, and the apodized resolution is 6 cm$^{-1}$. The recording time is 20 µs, identical to the spectrum shown in Fig. 3a. The signal-to-noise ratio of the most intense toluene band at 1001 cm$^{-1}$ is 1020. Therefore, for the same measurement time, the spectrum recorded with the 1-GHz dual-comb system (Fig. 3a) has a signal-to-noise ratio about one order of magnitude worse than the spectrum measured with the 100-MHz system, whereas the pulse energy is eight-fold weaker and therefore the CARS signal five-hundred-and-twelve times fainter. Such favorable scaling may be explained by several factors. The number of pump-probe sequences during the measurement time is 10 times higher with the 1-GHz system than with that of 100 MHz. The heterodyne detection by mixing with the local oscillator field of the spectrally overlapping probe laser beam benefits the signal-to-noise ratio.

Such results contribute to a current effort to increase the acquisition speeds and rates of techniques of broadband nonlinear Raman spectroscopy (see e.g. [24-28]). Moreover, direct frequency comb spectroscopy of samples in the condensed phase attracts growing interest [7-12]. Such experiments will benefit [29] from frequency combs of large line spacing and might trigger the development of dedicated laser sources. While linear-absorption dual-comb spectroscopy [9-11] may straightforwardly take advantage of such combs of high repetition frequency, nonlinear dual-comb spectroscopy of transitions of short coherence times still calls for optimized strategies. In this letter, we demonstrate that mode-locked lasers of 1-GHz repetition frequency can be successfully harnessed for coherent Raman dual-comb spectroscopy of neat liquids. We experimentally show duty-cycles that are ten-fold better than with lasers of a repetition frequency of 100 MHz [7], as well as refresh times that are improved twenty fold. The span of the 1-GHz lasers, of 1500 cm$^{-1}$, allows in principle refresh rates $\delta f_{rep}$ up to about 10 kHz but improvements to the signal-to-noise ratio will then be required. With such future enhancements to the signal-to-noise ratio and with the recent availability of commercial Ti:Sa femtosecond oscillators of 10-GHz repetition frequency, the full potential of dual-comb coherent Raman spectroscopy without moving parts may be realized [30]. Broadband dual-comb Raman spectra measured on a microsecond-scale at a refresh rate of several tens of kHz may become within reach and dual-comb CARS might evolve into a powerful tool for studies of short-lived transient species, for high sensitivity experiments requiring rapid signal averaging, or for fast hyper-spectral imaging.

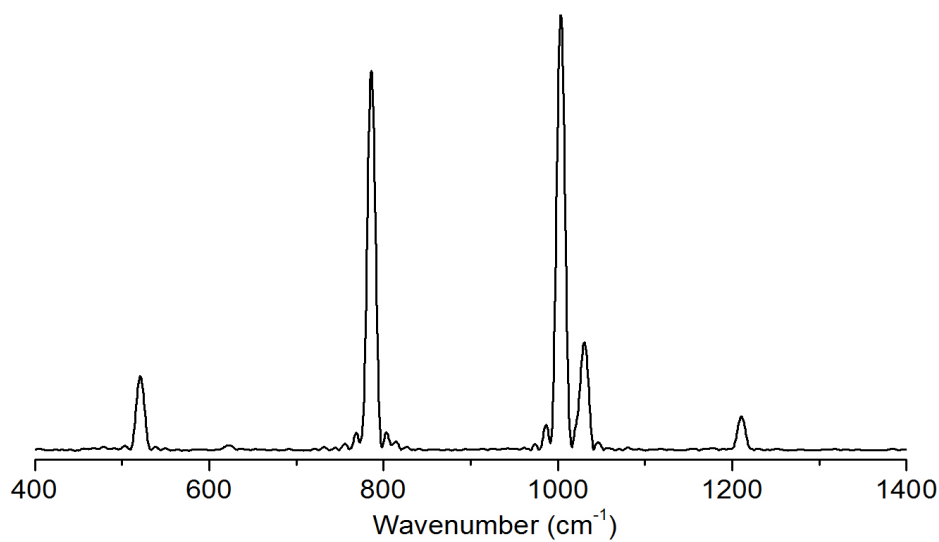

**Figure 6:** Experimental dual-comb spectrum of toluene, measured with two 100-MHz femtosecond lasers .

**Acknowledgements.** Financial support was provided by the European Research Council (Advanced Investigator Grant FP7-ERC-Multicomb no. 267854), the Max Planck Foundation and Munich Center for Advanced Photonics. We thank Dr. S. Holzner and A. Hipke for fruitful discussions and experimental advice.